\pdfoutput=1 
\documentclass[journal]{IEEEtran}

\usepackage{cite}

\ifCLASSINFOpdf
    \usepackage[pdftex]{graphicx}
\else
    \usepackage[dvips]{graphicx}
\fi

\usepackage{amsmath}

\begin{document}

\title{Reflectionless Filter Topologies Supporting Arbitrary Ladder Prototypes}
%
%
%

\author{Matthew~A.~Morgan,~\IEEEmembership{Senior Member,~IEEE,}
        Wavley~M.~Groves~III,~\IEEEmembership{Member,~IEEE,}
        and Tod~A.~Boyd%
\thanks{Manuscript received???

M. Morgan, W. Groves, and T. Boyd are with the Central Development Laboratory, National Radio Astronomy Observatory, Charlottesville,
VA, 22903 USA (e-mail: matt.morgan@nrao.edu). The National Radio Astronomy Observatory is a facility of the National Science Foundation operated under cooperative agreement by Asoociated Universities, Inc.}}

\markboth{IEEE Transactions on Circuits and Systems-I,~Vol.~xx, No.~x, xxx~2018}%
{Morgan \MakeLowercase{\textit{et al.}}: Reflectionless Filter Topologies Supporting Arbitrary Ladder Prototypes}

\IEEEpubid{0000--0000/00\$00.00~\copyright~2018 IEEE}


\maketitle

\begin{abstract}
Modifications of the authors' previously-published, generalized, lumped-element, reflectionless filter topologies are presented which remove the original constraints on the relative values of its prototype parameters. Thus, any transfer function which can be realized as the transmission or reflection coefficient of a conventional ladder prototype may now be implemented in reflectionless form --- that is, having the same transfer function in transmission but with identically zero reflection coefficient at both ports and at all frequencies from DC to infinity, given ideal elements. The theoretical basis of these modifications is explained, and then tested via the construction of passive, reflectionless low-pass filter prototypes that, in the prior topology, would have required negative reactive elements. The results show excellent agreement with theory.
\end{abstract}


\section{Introduction}

\IEEEPARstart{C}{onventional} electronic filters reject unwanted spectral content by reflection. Made with only lossless elements, their input impedance becomes almost purely reactive in the stop-band, often with rapidly varying phase as a function of frequency. Such reactive terminations are well-known to interact badly with broadband and/or nonlinear components commonly connected to filters in electronic systems, such as mixers, multipliers, amplifiers, and Analog-to-Digital Converters (ADCs). Mixers may exhibit enhanced generation of spurious products \cite{AN-75-007}. Mixers and multipliers will almost certainly undergo rapid changes in conversion efficiency as a consequence of resonances at harmonics or idler frequencies. Amplifiers may have degraded dynamic range and even become unstable. Finally, an ADC will invariably generate switching transients that may become trapped between itself and the neighboring anti-aliasing filter, ensuring the very presence of aliased harmonics that the filter was intended to eliminate.

\begin{figure}[!t]
    \centering
    \includegraphics{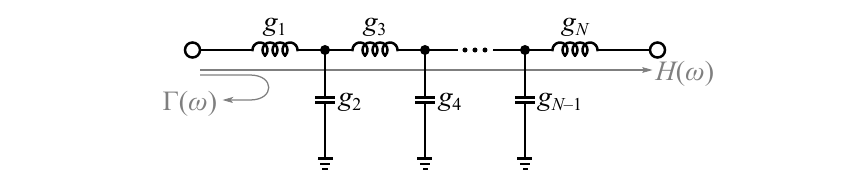}\\
    (a)\\
    \includegraphics{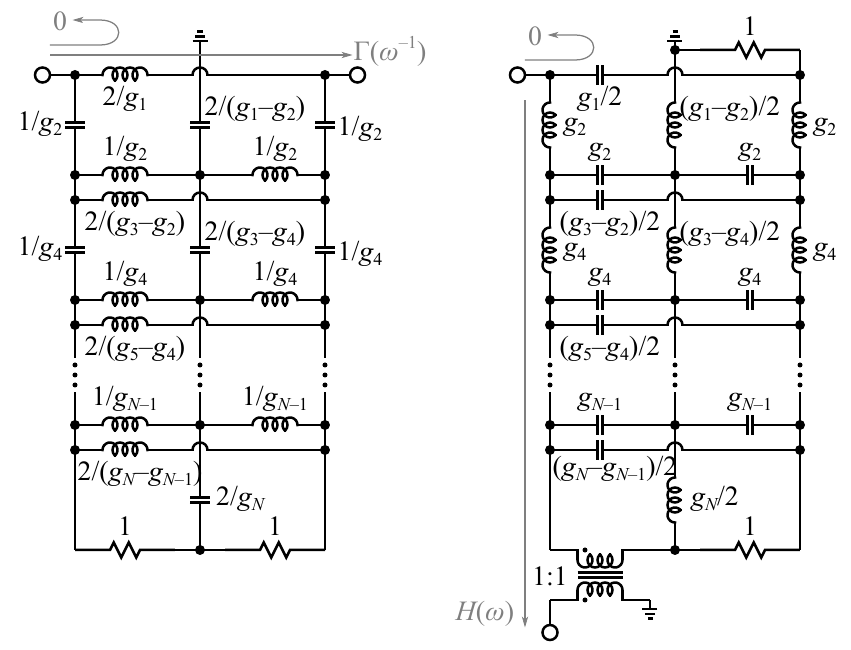}\\
    \hspace{0.1in}(b)\hspace{1.7in}(c)
    \caption{(a) Conventional ladder-topology, low-pass filter prototype having reflection response $\Gamma(\omega)$ and transmission response $H(\omega)$. (b) Generalized reflectionless filter topology having transmission response equal to $\Gamma\left(\omega^{-1}\right)$, and (c) reflectionless filter topology having transmission response equal to $H(\omega)$. The elements in all cases are labeled with normalized quantities.}
    \label{fig:original_topo}
\end{figure}

These issues have motivated a resurgence of research into absorptive filter techniques in recent years \cite{khalajamirhosseini2017,psychogiou2017,chien2017}. In particular, the authors previously developed symmetric, reflectionless filter topologies, which exhibit constant, purely resistive impedance-matching at all frequencies --- pass-band, stop-band, and transition-band --- from either port, given the equivalent termination on the opposite port \cite{morgan_theoretical,morgan_structures,morgan8392495,morgan9705467,morgan9923540,morgan_optimal, morgan_deep}. The culmination of that work resulted in the generalized topologies shown in Fig.~\ref{fig:original_topo}(b) and (c),
first published in this most general form in \cite{morgan_artech}. The labels represent element values that have been normalized for both cutoff frequency and impedance --- thus, the inductors may be assumed to be labeled in Henrys, the capacitors in Farads, and the resistors in Ohms, for a filter with cutoff frequency of 1 rad/s.

\IEEEpubidadjcol

The basis of these topologies is the conventional ladder prototype in Fig.~\ref{fig:original_topo}(a), having transmission response $H(\omega)$ and reflection response $\Gamma(\omega)$. The prototype parameters, $g_1\ldots g_N$, may be found tabulated for a variety of canonical responses (e.g. Chebyshev, Butterworth, etc.) in almost any textbook on electronic circuits.

The topology in Fig.~\ref{fig:original_topo}(b) is symmetric, and as such may be understood using even-/odd-mode analysis. The even-mode equivalent circuit (having simultaneous, identical inputs on both ports, and thus incurring an open-circuit boundary condition on the symmetry plane) reduces to a simple, terminated high-pass ladder with elements $1/g_1$, $1/g_2$, $1/g_3$, etc. --- in other words, a high-pass transformation of Fig.\ref{fig:original_topo}(a). The odd-mode equivalent circuit (having simultaneous, identical inputs with opposite signs on the ports, and thus incurring a short-circuit boundary condition on the symmetry plane) reduces to the dual high-pass ladder, starting with a shunt inductor instead of a series capacitor. The duality of these two equivalent circuits is sufficient to guarantee that the input impedance of the complete filter is constant, and given by the normalizing, or characteristic impedance \cite{morgan_structures}. The transmission response is simply given by the reflection coefficient of the even-mode equivalent circuit, or equivalently the reflection from a terminated, high-pass ladder based on the same list of prototype parameters. For example, if the parameters $g_1$, $g_2$, $g_3$, etc. are taken from the tables found in numerous textbooks for Chebyshev filters, than the transmission response of the reflectionless filter shown will be that of an \emph{inverse} Chebyshev, or Chebyshev type II filter.

The circuit in Fig.~\ref{fig:original_topo}(c) results from a high-pass transformation of the circuit in Fig.~\ref{fig:original_topo}(b), where the the original output port is terminated and instead the output is taken differentially (using a balun transformer) from the position of the previous termination resistor on the input side. The constant-impedance --- or equivalently, reflectionless --- property is preserved, however the transmission response is now the same as that of Fig.~\ref{fig:original_topo}(a) (e.g. a Chebyshev type I filter, if said prototype parameters are used).

\section{Negative Element Mitigation}\label{sec:negative_element_mitigation}

The reflectionless topologies in Fig.~\ref{fig:original_topo} have one major shortcoming, however, which is that the odd-numbered prototype parameters must be greater than or equal to the adjacent even-numbered parameters, lest some of the elements become negative --- such as the grounded capacitor in Fig.~\ref{fig:original_topo}(b) labeled $2/(g_1-g_2)$. In the case of a Chebyshev filter, this constrains the ripple factor, $\varepsilon$, for any order $N$ such that
\begin{equation}\label{eq:ripple_factor}
    \varepsilon \ge \sqrt{e^{4\tanh^{-1}\left(e^{-\beta}\right)}}
\end{equation}
where
\begin{equation}
    \beta = \sqrt{\tfrac{1}{2}\sin\left(\tfrac{\pi}{N}\right)\tan\left(\tfrac{\pi}{N}\right)}
\end{equation}
Butterworth filters (which may be considered a limiting case of the Chebyshev filter as $\varepsilon\rightarrow 0$) are entirely excluded by this constraint.

\subsection{Negative Element Identities}

\begin{figure}[!t]
    \centering
    \includegraphics{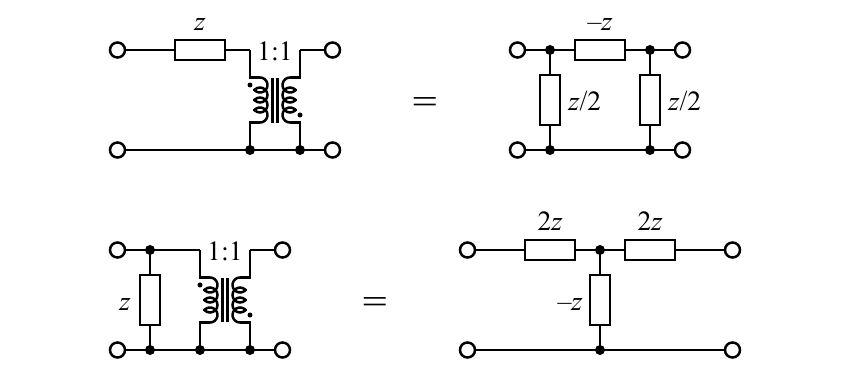}
    \caption{Two transformer circuit identities which are useful in eliminating negative elements. Note the dot convention signifies a current-reversing orientation.}
    \label{fig:identities}
\end{figure}

The key to solving this problem lies in recognizing that the negative elements do not appear in isolation, but rather are embedded in the filter with other elements that render the ensemble passive. It is thus possible to substitute an equivalent, all positive-element subcircuit for these groupings which preserves the behavior of the overall filter. Consider the two circuit identities shown in Fig.~\ref{fig:identities}.
In each case, a 1:1 transformer is wired in a current-inverting configuration (equivalent to a --1:1 turns ratio) and cascaded with a single impedance element, either series or parallel, and equates to a transformerless tee- or pi-network where the central element is negative. Take, for instance, the first circuit having an element in series with the transformer. The ABCD parameters of the cascade are given by
\begin{equation}
    \begin{pmatrix}A & B\\C & D\end{pmatrix} = \begin{pmatrix}1 & z\\0 & 1\end{pmatrix}\begin{pmatrix}-1 & 0\\0 & -1\end{pmatrix} = \begin{pmatrix}-1 & -z\\0 & -1\end{pmatrix}
\end{equation}
Converting this to an admittance matrix, we have
\begin{equation}
    \mathbf{Y} = \frac{1}{B}\begin{pmatrix}D & BC-AD\\-1 & A\end{pmatrix} = \begin{pmatrix}z^{-1} & z^{-1}\\z^{-1} & z^{-1}\end{pmatrix}
\end{equation}
which is easily recognized as a pi-network with $z/2$ in the shunt branches and $-z$ in the series branch. Similarly, the ABCD parameters of the shunt element in cascade with the inverting transformer are
\begin{equation}
    \begin{pmatrix}A & B\\C & D\end{pmatrix} = \begin{pmatrix}1 & 0\\z^{-1} & 1\end{pmatrix}\begin{pmatrix}-1 & 0\\0 & -1\end{pmatrix} = \begin{pmatrix}-1 & 0\\-z^{-1} & -1\end{pmatrix}
\end{equation}
and the equivalent impedance matrix is
\begin{equation}
    \mathbf{Z} = \frac{1}{C}\begin{pmatrix}A & AD-BC\\1 & D\end{pmatrix} = \begin{pmatrix}z & -z\\-z & z\end{pmatrix}
\end{equation}
which is clearly equivalent to a tee-network having a parallel element, $-z$, with series $2z$ elements.

\subsection{Parallel Branch Replacement}

\begin{figure}[!t]
    \centering
    \includegraphics{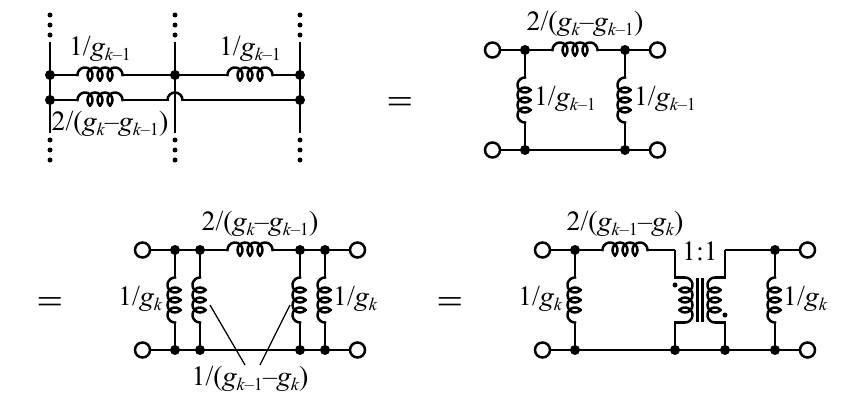}
    \caption{Sequence of steps required to eliminate the negative element in a three-inductor ring ($g_k<g_{k-1}$).}
    \label{fig:parallel_branch}
\end{figure}

The first of these identities is useful for getting rid of the negative elements which appear in a three-element ring (comprising inductors in the low-pass prototype of Fig.~\ref{fig:original_topo}(b)). Since the elements surrounding the negative one in the ring do not have exactly the right value to apply the identity directly, we must first break them down into parallel components, as indicated in Fig.~\ref{fig:parallel_branch}.
For this example, we assume that $g_k<g_{k-1}$, causing the element in the original topology labeled $2/\left(g_k-g_{k-1}\right)$ to be negative. After the identity is applied, we are left with only positive elements.

\subsection{Series Branch Replacement}

The other place in which a negative element may appear in our original topologies is in one of the series branches along the center line, such as the grounded capacitor labeled $2/\left(g_1-g_2\right)$ in Fig.~\ref{fig:original_topo}(b), where we assume $g_1<g_2$. Application of the identities in Fig.~\ref{fig:identities} is made more difficult this time by the fact that the negative element and its stabilizing counterparts --- the $1/g_2$ capacitors to the left and right of the center line --- are not even directly connected to each other.

\begin{figure}[!t]
    \centering
    \includegraphics{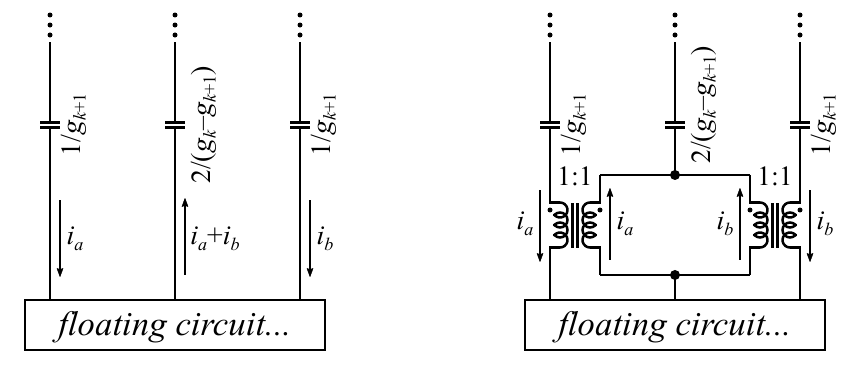}\\
    (a)\hspace{1.8in}(b)\\
    \vspace{3ex}
    \includegraphics{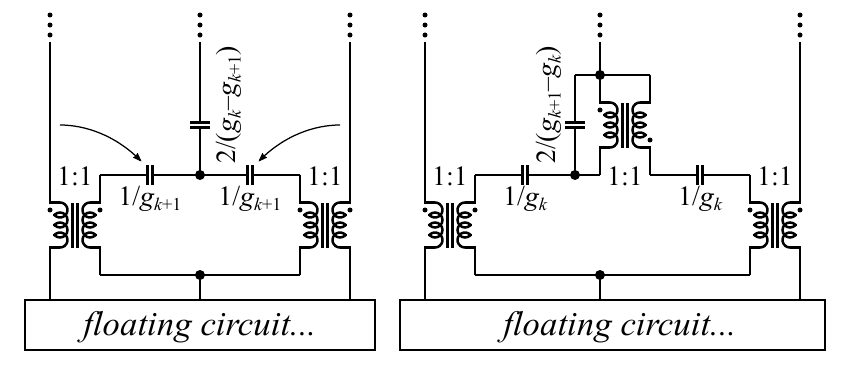}\\
    (c)\hspace{1.5in}(d)\hspace{0.1in}\phantom{.}\\
    \vspace{3ex}
    \includegraphics{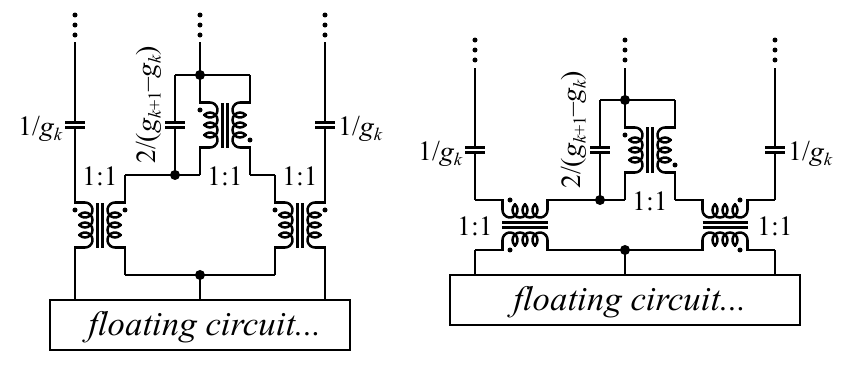}\\
    (e)\hspace{1.6in}(f)\hspace{0.2in}
    \caption{Sequence of steps required to eliminate the negative element in three series-capacitor branches, when $g_k<g_{k+1}$. (a) Illustration of current-summing in center branch. (b) Introduction of transformers to hold the current-summing condition during the following steps. (c) Transfer of outer capacitors to the inside. (d) Application of inverting-transformer identity. (e) Optional relocation of the residual series capacitors back to the outer branches. (f) Optional rotation of transformers to improve low-frequency isolation.}
    \label{fig:series_branch}
\end{figure}

To resolve this, we note that there is no ground sink anywhere in the circuit below these series branches. Thus, any current which flows up through the center line must be balanced by currents flowing down through the left and right branches, as illustrated in Fig.~\ref{fig:series_branch}(a).
We may thus introduce a pair of transformers, as in Fig.~\ref{fig:series_branch}(b), to hold this condition when we make topological changes in the next steps. In Fig.~\ref{fig:series_branch}(c), we move the outer two capacitors from one side of the adjoining transformer to the other, thus collecting all the capacitors into a tee-network in the middle. That, then, allows us to apply the second transformer-identity from Fig.~\ref{fig:identities}. As before, the element values do not have the necessary impedance ratio, so we must first split the capacitors into series elements (not shown). The result, after application of the identity, is given in Fig.~\ref{fig:series_branch}(d).

In principle, we could stop here, as we now have an all-positive equivalent circuit. Some optional modifications, however, may be made to further improve performance in practical cases where the transformers do not function down to zero frequency. In circuits such as Fig.~\ref{fig:original_topo}(b), the floating part of the circuit below the series capacitors is active only in the stop-band. To improve the isolation of this part of the circuit in the stop-band, despite degradation of the transformer action in this regime, we may move the residual series capacitors back to the outer branches, as indicated in Fig.~\ref{fig:series_branch}(e). We may also rotate the transformers as shown in Fig.~\ref{fig:series_branch}(f), which has no effect in the ideal case but provides better isolation given realistic transformers at low frequency.

Note that these last two changes make the most sense when the circuit is low-pass, and conforms to the style in Fig.~\ref{fig:original_topo}(b). If a high-pass transformation is applied, or if the style in Fig.~\ref{fig:original_topo}(c) is used, it may be best to omit them.

\section{Specific Filters and Experimental Results}

Which particular groupings of elements will require the substitutions outlined in Section~\ref{sec:negative_element_mitigation} will depend, of course, on the values of the prototype parameters, which in turn depend on the particular filter response that is desired. The technique for negative-element mitigation presented in this paper has made possible a much wider variety of responses than we could previously access in a totally reflectionless topology. Several of those newly achievable responses will now be discussed. Test filters which were constructed to verify the theory will also be presented.

\subsection{Deep-Rejection Chebyshev Type II Filters}

For Chebyshev filters just beyond the ripple factor limit in \eqref{eq:ripple_factor}, we have $g_1<g_2$ and, by symmetry, $g_N<g_{N-1}$, but everywhere else $g_k>g_{k\pm 1}$ where $k$ is odd. Thus, only the first capacitor group and the last inductor group in the odd-order low-pass prototype need these substitutions.

\begin{figure}[!t]
    \centering
    \includegraphics{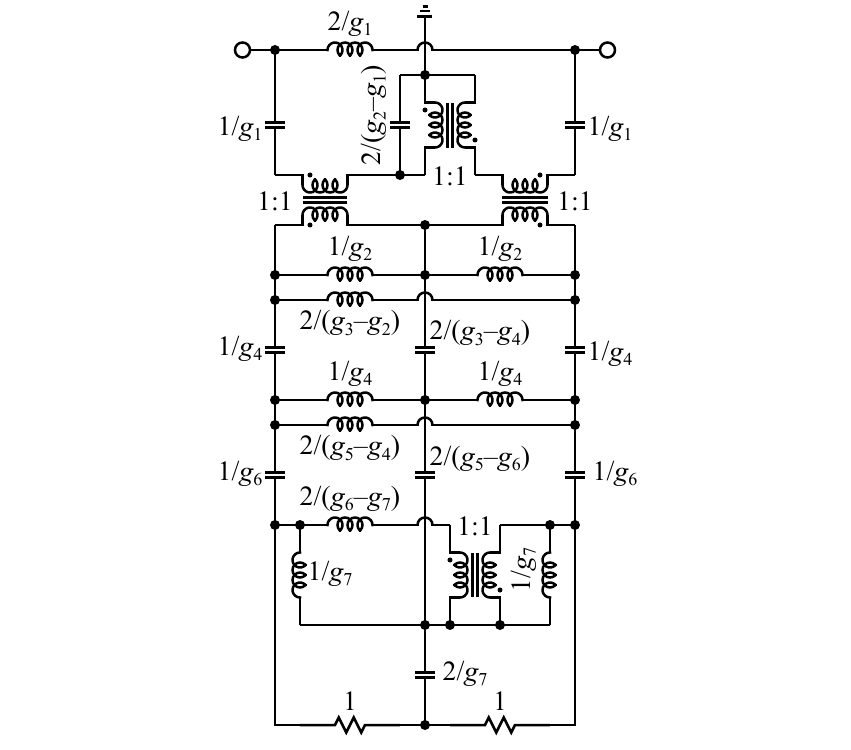}
    \caption{Topology for a seventh-order, low-pass, reflectionless Chebyshev type II filter with ripple factor less than the limit given in \eqref{eq:ripple_factor}.}
    \label{fig:chebII7}
\end{figure}

\begin{figure}[!t]
    \centering
    \includegraphics{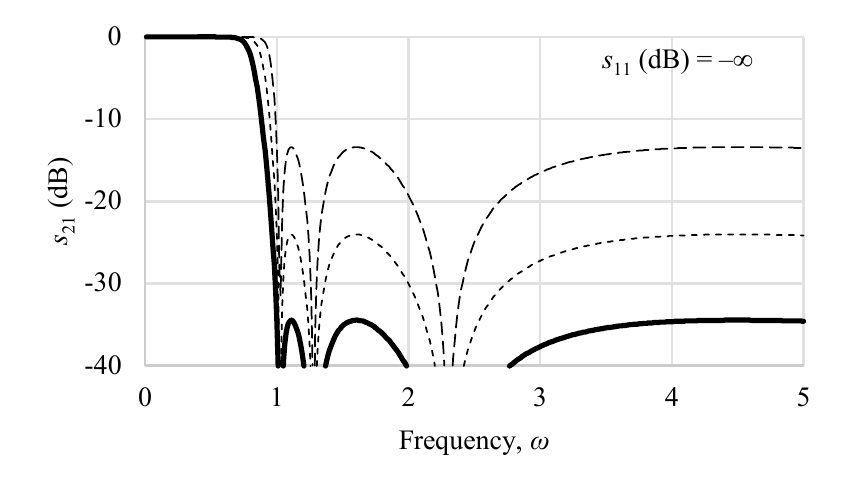}
    \caption{Normalized frequency response of the reflectionless, Chebyshev type II filter topology in Fig.~\ref{fig:chebII7}, where the ripple factor varies from $\varepsilon = 0.2187$ to $\varepsilon=0.01891$.}
    \label{fig:chebII7_plot}
\end{figure}

The final topology for a seventh-order, low-pass, reflectionless Chebyshev type II filter with ripple factor lower than the limit in \eqref{eq:ripple_factor} is thus shown in Fig.~\ref{fig:chebII7}.
The theoretical response of such a filter is given in Fig.~\ref{fig:chebII7_plot}
for several ripple factors.

One may discover that as the ripple factor for this filter decreases below $\varepsilon \approx 0.2187$ (the previous limit) the next pair of adjacent prototype parameters, $g_3$ and $g_4$, approach one another in value. At $\varepsilon \approx 0.08191$, they become equal, and below that threshold new negative elements appear in our topology. This can be addressed with additional substitutions of our equivalent subcircuits, however the stop-band rejection for a type II filter at this point is approximately 34.5 dB, which is sufficient for many applications. It is convenient, in fact, to realize a reflectionless filter at this specific ripple factor, since two of the elements become infinite and therefore vanish --- the infinite capacitor, $2/\left(g_3-g_4\right)$, being replaced by a short-circuit, and the infinite inductor, $2/\left(g_5-g_4\right)$, being replaced by an open circuit.

\begin{figure}[!t]
    \centering
    \includegraphics{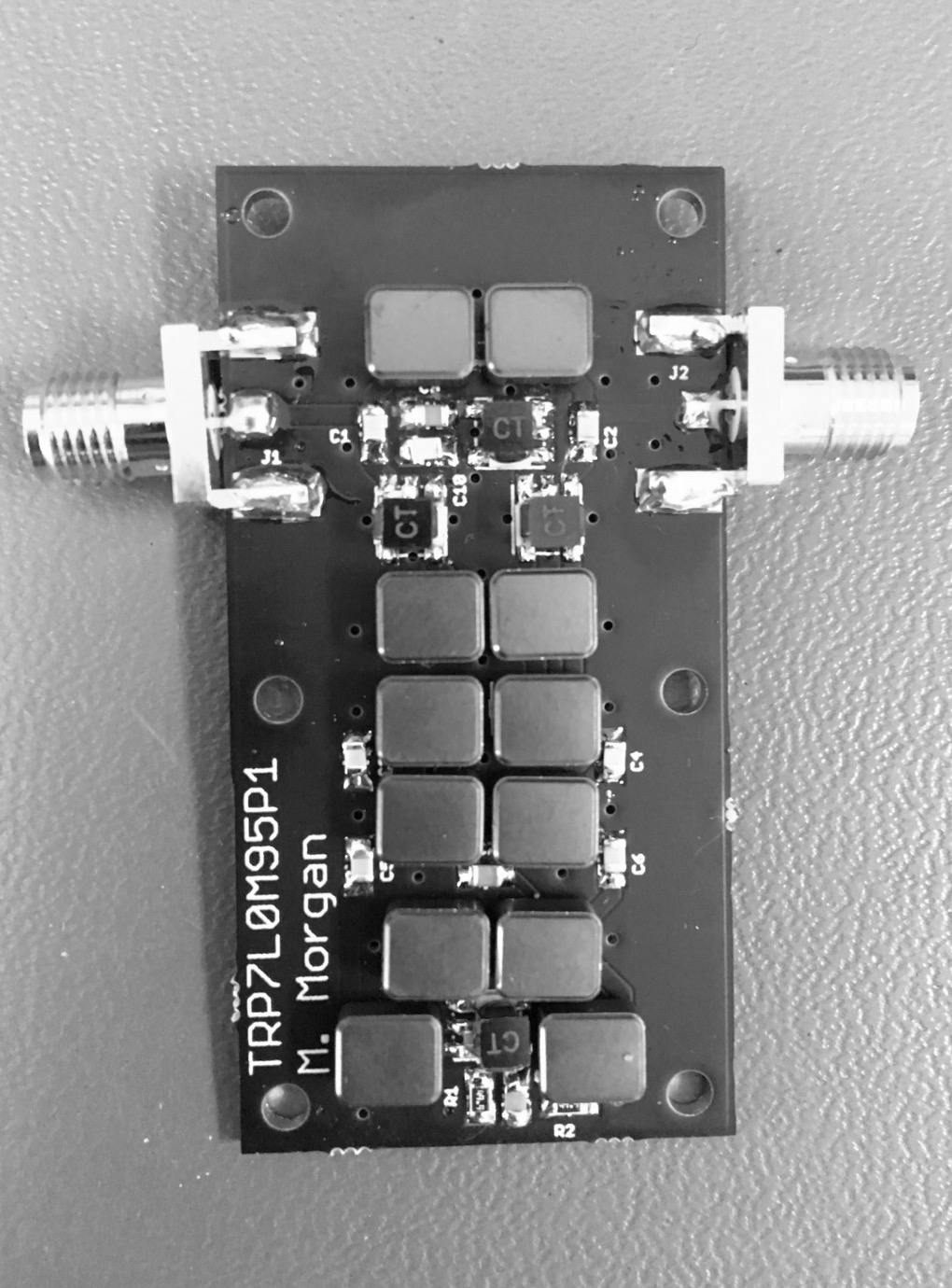}
    \caption{Photo of reflectionless, deep-rejection, Chebyshev type II filter.}
    \label{fig:trp7l0m95p1}
\end{figure}

This is the approach taken for the prototype test circuit presented in Fig.~\ref{fig:trp7l0m95p1}.
For simplicity, we have implemented this filter at a fairly modest cutoff frequency, with its stop-band corner at about 1 MHz, where discrete, surface-mount lumped elements and quality transformers are most readily available, but scaling to higher frequencies is straightforward so long as the needed elements can be fabricated. Further, in keeping with filter theory traditions, we have limited our initial exploration to a low-pass design, but high-pass, band-pass, and band-stop responses are clearly obtainable using the well-known frequency transformations.

The inductors utilized were the 1812FS series from Coilcraft, with 5\% tolerance, the capacitors were KEMET's C0G ceramic capacitors with 1\% tolerance, and the transformers were model TC1-6X+ from Mini-Circuits. In some cases, inductors from the original topology were split into series pairs so that the desired values may be more easily realized using the available parts.

\begin{figure}[!t]
    \centering
    \includegraphics{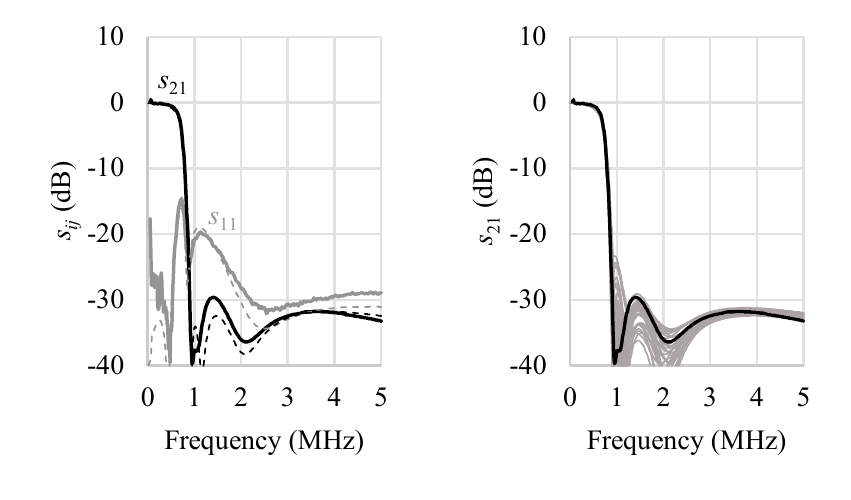}\\
    \hspace{0.4in}(a)\hspace{1.5in}(b)
    \caption{(a) Measured performance (solid lines) of the deep-rejection Chebyshev type II filter compared with the model incorporating nominal parasitics (dashed lines). (b) Yield analysis showing expected variations with element tolerance.}
    \label{fig:trp7l0m95p1_plots}
\end{figure}

The response of this filter is shown in Fig.~\ref{fig:trp7l0m95p1_plots}(a),
along with the modeled results (dashed lines) after incorporating the predicted nominal parasitics. Note that the reflection coefficient is better than --15 dB in the transition band and is nearly --30 dB for most of the stop-band. The departure from the modeled return loss below 500 kHz is believed to be an artifact of the measurement, as the vector network analyzer (Hewlett Packard model 8753D) loses sensitivity at the bottom end of its frequency range.

As a seventh-order design, the low-pass Chebyshev type II response should nominally have three finite-frequency transmission zeros, as indicated by the dashed line in Fig.~\ref{fig:trp7l0m95p1_plots}(a), but one of these has nearly disappeared in the measured $s_{21}$ curve. A sensitivity analysis based on the advertised tolerance of the elements, shown in Fig.~\ref{fig:trp7l0m95p1_plots}, confirms that this is within the expected range of random variations for this implementation.

\subsection{Butterworth Filters}

\begin{figure}[!t]
    \centering
    \includegraphics{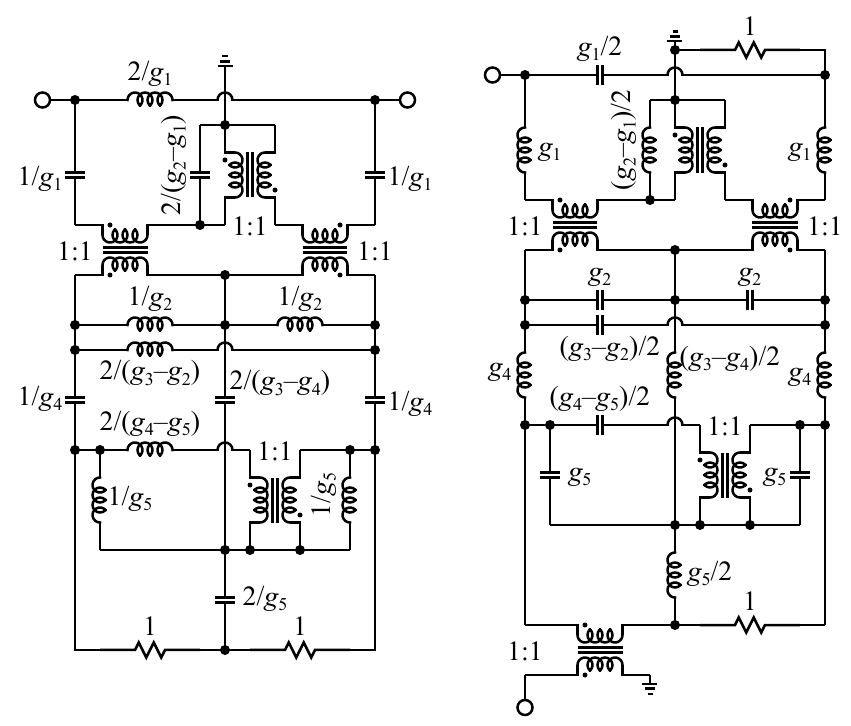}\\
    \hspace{0.2in}(a)\hspace{1.6in}(b)
    \caption{Two possible topologies for a fifth-order, reflectionless Butterworth filter.}
    \label{fig:butterworth}
\end{figure}

As the ripple factor approaches zero, the Chebyshev response (whether type I or Type II) approaches that of a maximally flat, or Butterworth filter, valued for its monotonic response and smooth group delay. Interestingly, this admits two different topological solutions for the same reflectionless response --- that based on the reflection of a ladder prototype, as in Fig.~\ref{fig:original_topo}(b), and that based on the transmission of a ladder prototype, as in Fig.~\ref{fig:original_topo}(c). The final topologies for a fifth-order Butterworth filter, after the required substitutions to eliminate negative elements, are shown in Fig.~\ref{fig:butterworth}.

Though equivalent in principle, these two topologies are hardly equivalent in practice. As a monotonic filter, the stop-band rejection of the Butterworth is designed to increase without bound as you move further away from cutoff. For the topology in Fig.~\ref{fig:butterworth}(a), having only one series element between the two ports, the highest levels of attenuation can only be achieved if the remainder of the circuit shunts the two ports to ground and cancels out the series element with a carefully tuned negative susceptance. The latter condition is especially sensitive to small element variations.

\begin{figure}[!t]
    \centering
    \includegraphics{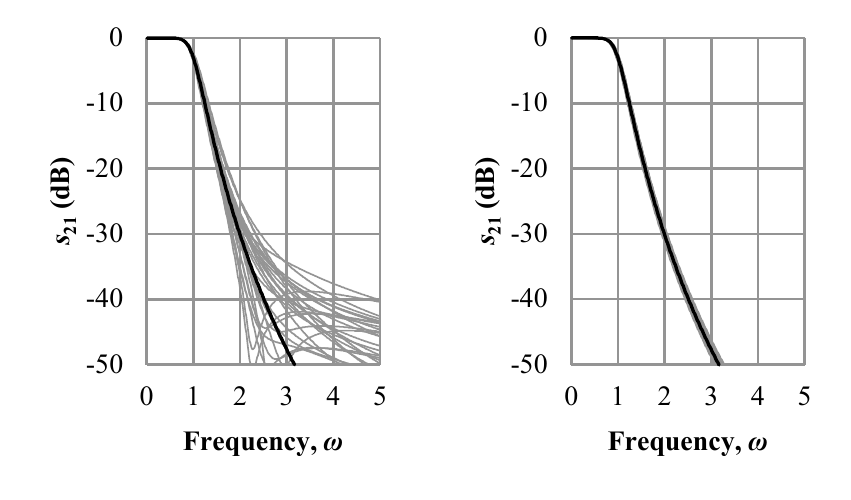}\\
    \hspace{0.3in}(a)\hspace{1.6in}(b)
    \caption{Yield analysis applied to (a) topology in Fig.\ref{fig:butterworth}(a) and (b) topology in Fig.\ref{fig:butterworth}(b), with 5\% standard deviation element variations.}
    \label{fig:butt_sim}
\end{figure}

The topology in Fig.~\ref{fig:butterworth}(b), however, is much more robust, since attenuation is achieved by a cascade of many series and shunt elements, much like the original ladder prototype. Small errors do not have such an impact on the stop-band response. To better illustrate the difference, a yield analysis was performed on both topologies where the element values were allowed to vary with a 5\% standard deviation. The results, plotted in Fig.~\ref{fig:butt_sim},
are a striking illustration of the comparative sensitivity of these two models to lumped-element values. On the other hand, the common-mode rejection of the output transformer/balun in Fig.\ref{fig:butterworth}(b), not included in this analysis, is critical to  deep stop-band performance.

\begin{figure}[!tb]
    \centering
    \includegraphics{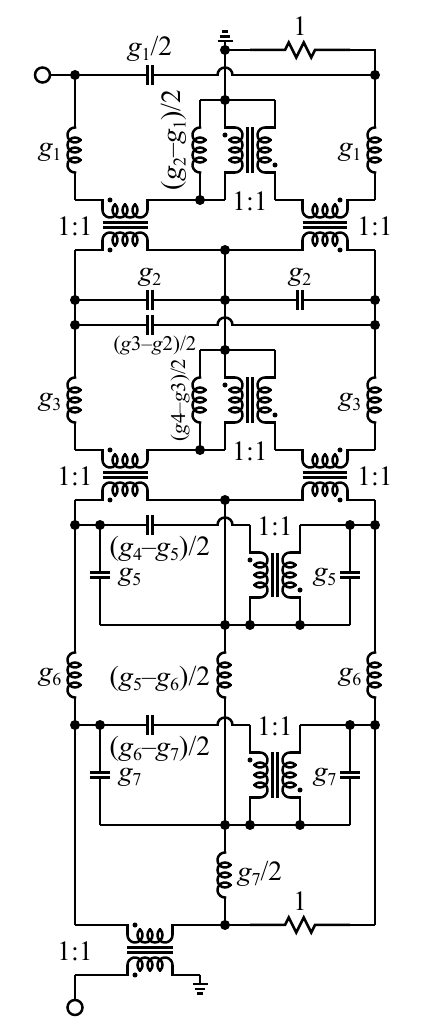}
    \includegraphics{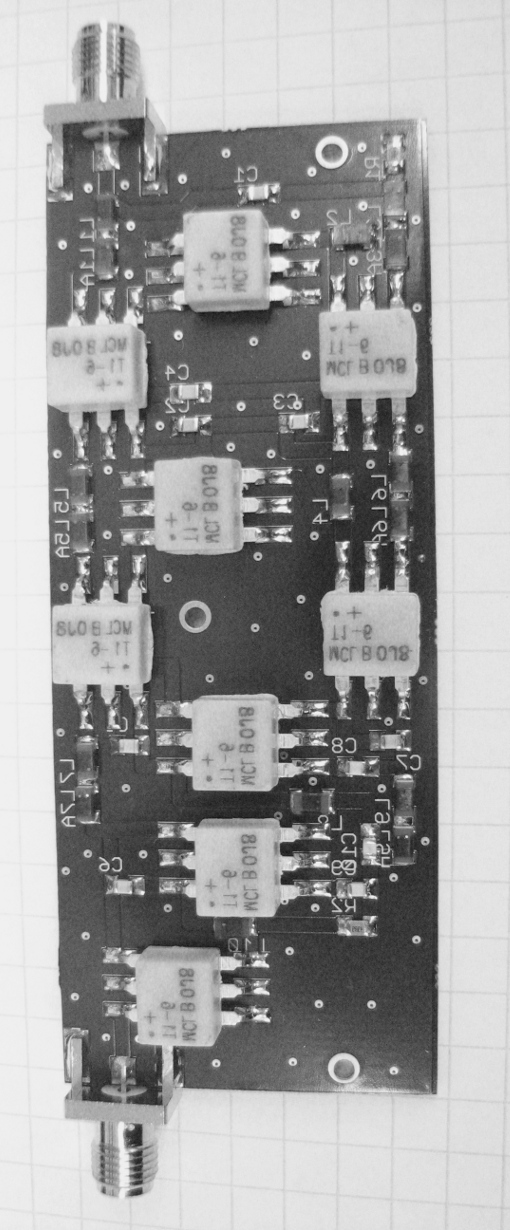}\\
    \hspace{0.1in}(a)\hspace{1.6in}(b)\\
    \caption{Seventh-order Butterworth reflectionless filter (a) topology and (b) test circuit board.}
    \label{fig:butt_test}
\end{figure}

The topology for a seventh-order Butterworth filter is given in Figure~\ref{fig:butt_test}(a).
A total of four negative-element identities have been applied, two of each type described in Section~\ref{sec:negative_element_mitigation}. A test circuit constructed using Coilcraft 1812CS inductors and Mini-Circuits T1-6+ transformers, this one tuned to 10 MHz corner frequency, is shown in Figure~\ref{fig:butt_test}(b).

\begin{figure}[!tb]
    \centering
    \includegraphics{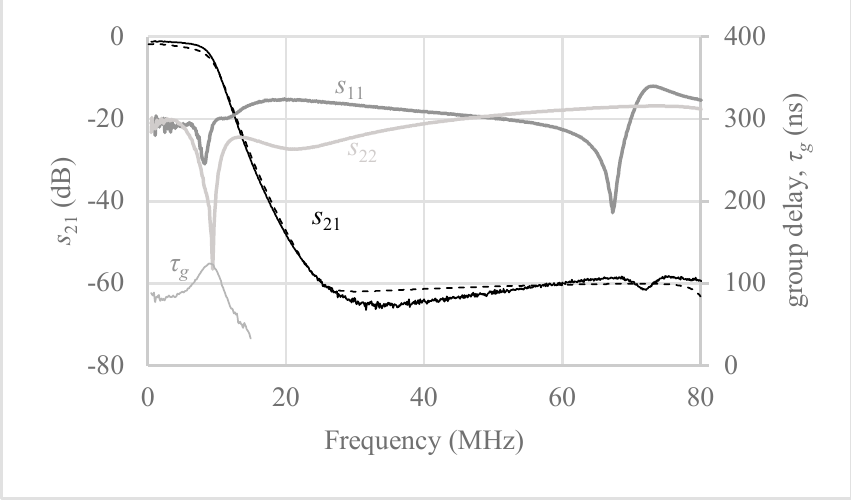}
    \caption{Measured scattering parameters and group delay of the reflectionless Butterworth filter. The modeled transmission coefficient is shown in a dashed line.}
    \label{fig:butt_meas}
\end{figure}

The measured performance is plotted in Figure~\ref{fig:butt_meas}.
Note that it achieves stop-band rejection of about 60 dB, with return loss better than 15 dB up to at least 70 MHz. Although noisy at this level, the 60 dB attenuation is not simply instrumental error, but rather it was predicted from the parasitics of the transformer used, as indicated by the dashed line. The pass-band group delay is also given in the lower left of the plot (but on the right-hand scale) to illustrate that the transmission response of this filter is exactly the same as what one would achieve with a conventional (reflective) Butterworth filter, including the desirable phase characteristics in this case.

\subsection{Legendre-Papoulis Filters}

Thus far, the canonical responses that we have examined had prototype parameter lists that were symmetric about the center. As such, the substitutions for negative element mitigation always came in pairs, one each of the two types described in Section~\ref{sec:negative_element_mitigation}. Now let us look at a response whose prototype parameter lists are not symmetric, namely the Legendre-Papoulis or Optimal-L filter \cite{papoulis1958, bond2004}. Like Butterworth, they are monotonic in the pass-band (but not maximally flat) and have the steepest transition for any monotonic filter.

\begin{figure}[!t]
    \centering
    \includegraphics{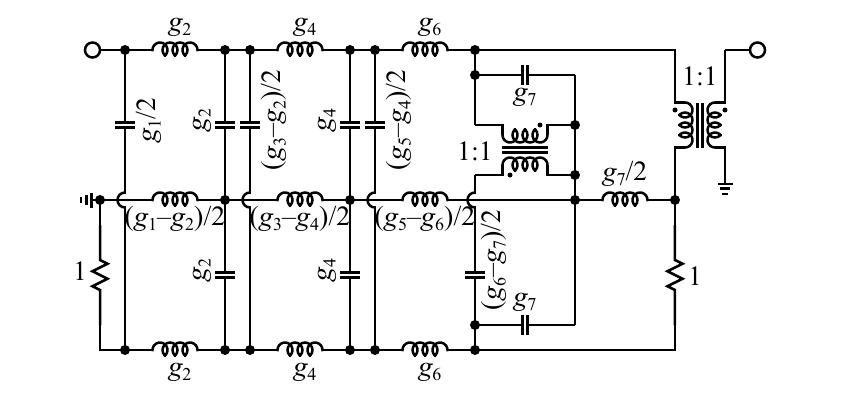}
    \caption{Reflectionless topology for a seventh-order Legendre-Papoulis filter.}
    \label{fig:legendre}
\end{figure}

The prototype parameters for Legendre filters of a few orders are given in Table~\ref{tab:legendre}.
\begin{table}[!t]
    \renewcommand{\arraystretch}{1.3}
    \caption{Prototype Parameters for Legendre-Papoulis Filters}
    \label{tab:legendre}
    \centering
    \begin{tabular}{@{}cccccccc@{}}
        \hline\hline
        $N$ & $g_1$ & $g_2$ & $g_3$ & $g_4$ & $g_5$ & $g_6$ & $g_7$\\
        \hline
        3 & 2.1801 & 1.3538 & 1.1737\\
        5 & 1.9990 & 1.5395 & 2.0672 & 1.4780 & 0.9512\\
        7 & 1.8640 & 1.5895 & 2.1506 & 1.7270 & 1.9394 & 1.4770 & 0.8394\\
        \hline\hline
    \end{tabular}
    \vspace{1ex}\\
    \emph{From}: C. R. Bond, 2004 \cite{bond2004}
\end{table}
In every case where $k$ is odd, we have $g_k>g_{k\pm1}$, except for $k=N$. Thus, one negative element appears in the basic reflectionless topology, and must be replaced with an all-positive equivalent. The resulting filter, where $N=7$ for illustration, is shown in Fig.~\ref{fig:legendre}.

\begin{figure}[!t]
    \centering
    \includegraphics{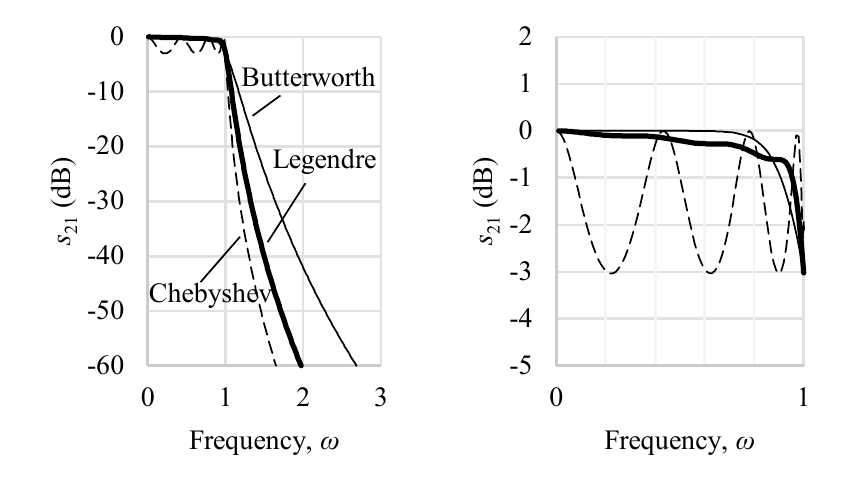}\\
    \hspace{0.3in}(a)\hspace{1.6in}(b)
    \caption{(a) Performance of reflectionless Legendre-Papoulis filter (thick line) compared against Butterworth (thin line) and Chebyshev type I (dashed line). (b) Detail of pass-bands.}
    \label{fig:legendre_plot}
\end{figure}

Legendre-Papoulis filters are considered a compromise between Chebyshev and Butterworth, having steeper transition than the latter while staying monotonic. A plot of the frequency response of our reflectionless Legendre-Papoulis filter using ideal elements is shown in Fig.~\ref{fig:legendre_plot}
with a thick line, compared to a Butterworth (thin line) and Chebyshev type I filter (dashed line). Note that the Legendre filter has nearly as steep a transition as the Chebyshev filter with 3 dB ripple, but is monotonic. On the other hand, as the pass-band detail in Fig.~\ref{fig:legendre_plot}(b) shows, it has uneven slope in the pass-band that is worse than the maximally flat Butterworth.

\begin{figure}[!tb]
    \centering
    \includegraphics{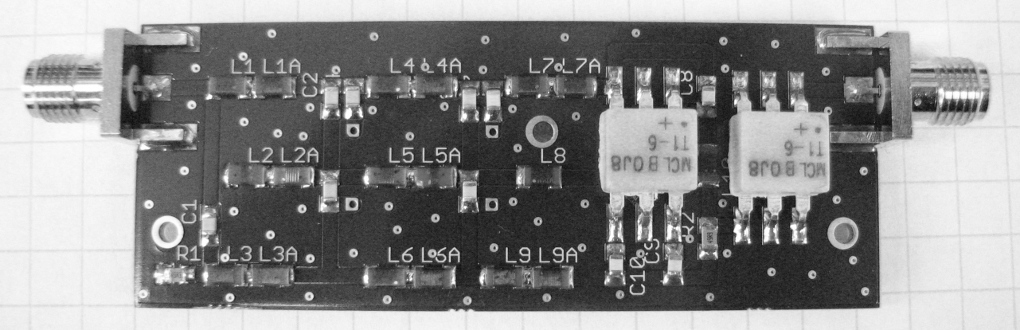}\\
    (a)\vspace{2ex}\\
    \includegraphics{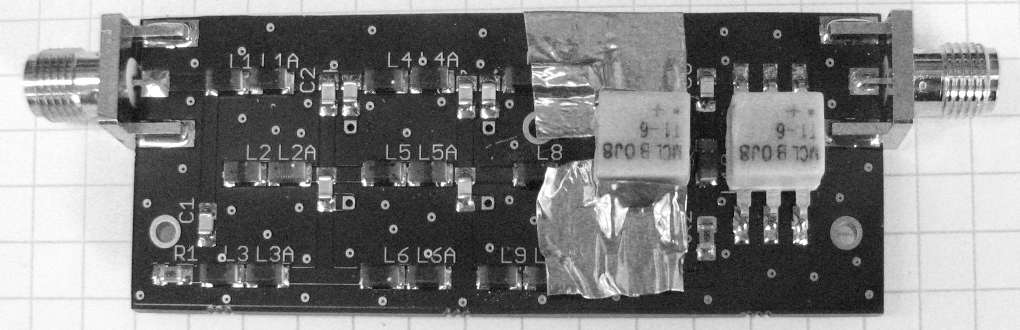}\\
    (b)
    \includegraphics{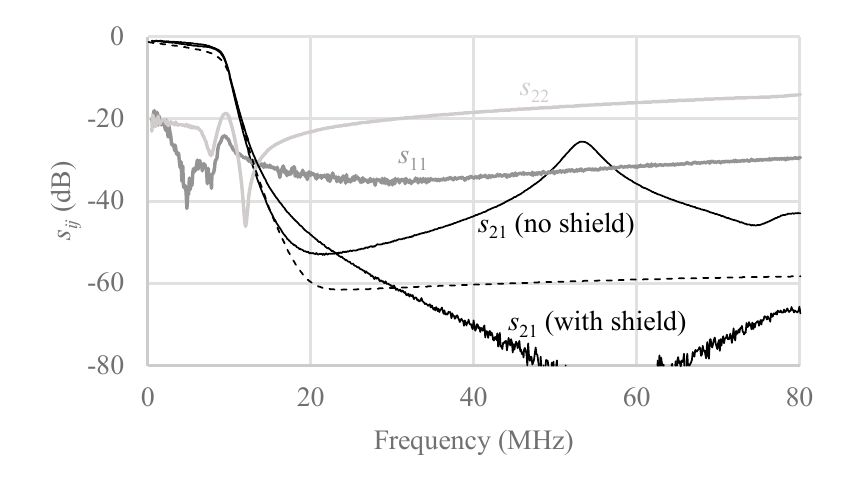}\\
    (c)
    \caption{(a) Prototype reflectionless Legendre-Papoulis filter without and (b) with electrical shielding around inverting transformer pins. (c) Measured performance. The modeled transmission coefficient is shown with a dashed line.}
    \label{fig:legend}
\end{figure}

To illustrate the practicality of this topology we have constructed a test circuit, also tuned to 10 MHz corner frequency like the Butterworth, and using similar parts. The circuit board is shown in Figure~\ref{fig:legend}(a).
The predicted stop-band attenuation was 60 dB, much like the Butterworth filter of Figure~\ref{fig:butt_meas}, but in this case a parasitic resonance at about 54 MHz disrupted the stop-band. Experimentation suggested that this could be caused by stray electrical coupling between the legs of the transformer package and adjacent components. To resolve the issue, copper tape was applied over the inverting transformer pins as shown in Figure~\ref{fig:legend}(b) (the thin adhesive layer on the copper tape was sufficient to prevent this from creating an electrical short). The measured performance with and without the copper tape is shown in Figure~\ref{fig:legend}(c).

\subsection{Even-Order Filters}

\begin{figure}[!t]
    \centering
    \includegraphics{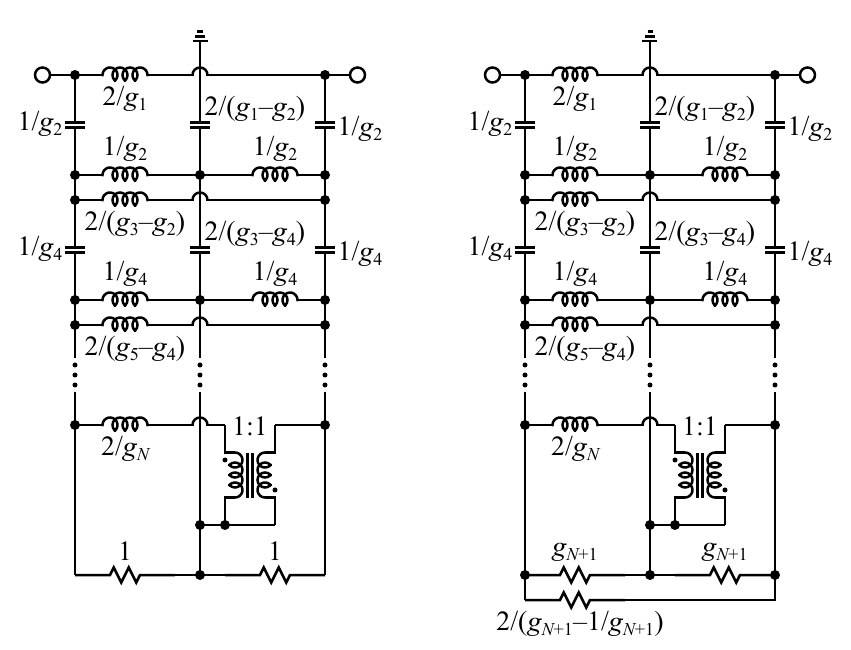}\\
    (a)\hspace{1.7in}(b)
    \caption{Even-order reflectionless filter topologies where the prototype ladder is terminated with a resistor of (a) unit normalized value and (b) non-unit normalized value.}
    \label{fig:even_order}
\end{figure}

Perhaps more of academic interest than practical is the fact that the mitigations presented in this paper also allow even-order responses to be achieved in reflectionless form, where only odd-order responses were possible before. Referring to the ladder prototype in Fig.~\ref{fig:original_topo}(a), an odd-order response may be realized by setting the final (odd-numbered) element to zero. This invariably leads to a negative difference in the final grouping of the original topology, which we now know how to mitigate. We thus arrive at the topology in Fig.~\ref{fig:even_order}(a),
where $N$ is now even. Of course, should any of the remaining elements of the filter become negative as a consequence of the relative prototype parameter values, the same substitutions described in the previous sections may be applied.

For many canonical responses, such as Chebyshev, the output port termination of the even-order ladder prototype does not have unit normalized value, having instead some other value which is conventionally labeled $g_{N+1}$. This requires a small modification of the topology, illustrated in Fig.~\ref{fig:even_order}(b), which allows the even- and odd-mode equivalent circuits to have dual-valued terminations. Note that the bottom-most resistor in this configuration could be negative if $g_{N+1}<1$, but the three resistors are arranged identically to the inductor sets above, and the same identity in Fig.~\ref{fig:identities} may be applied.

\begin{figure}[!tb]
    \centering
    \includegraphics{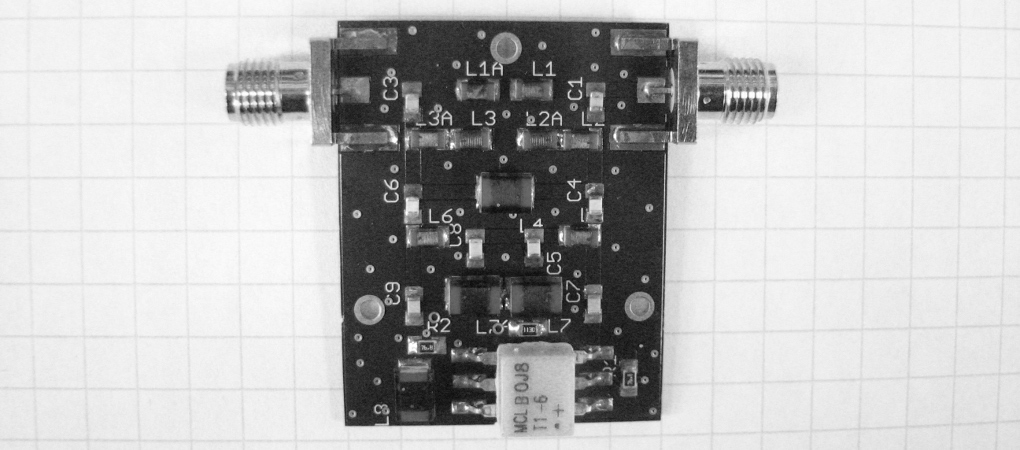}\\
    (a)\\
    \includegraphics{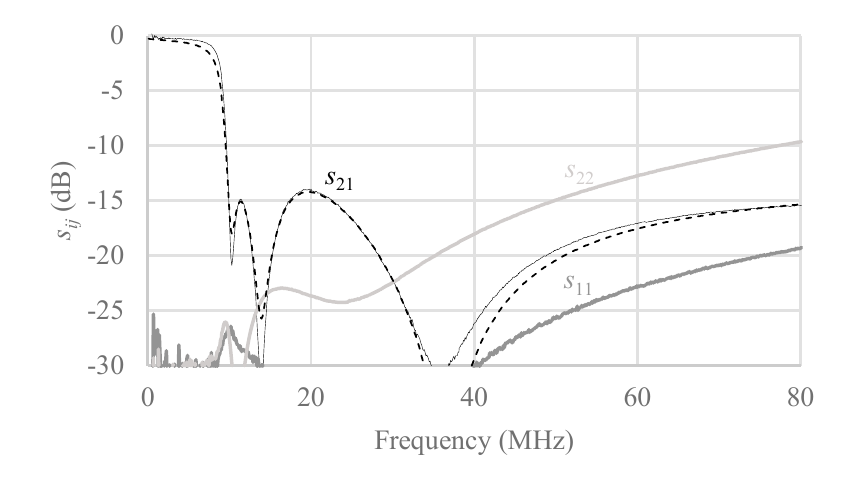}\\
    (b)
    \caption{(a) Prototype reflectionless Chebyshev type II filter of order $N=6$ and (b) its measured performance. The modeled transfer function is shown with a dashed line.}
    \label{fig:chebII6}
\end{figure}

Once again, to prove the theory, a test circuit was constructed and is shown in Figure~\ref{fig:chebII6}(a).
The measured results are shown in Figure~\ref{fig:chebII6}(b). Although similar on the surface to the seventh-order Chebyshev filter in Figure~\ref{fig:chebII7_plot}, its stop-band asymptotically approaches the ripple level at high frequencies instead of returning to a transmission-zero at infinity --- in practice, of course, neither will be the case as the response at very higher frequencies will succumb to the dominance of parasitics.

\section{Conclusion}
This paper has presented novel lumped-element filter topologies which are capable of implementing the same transfer functions as those that are achievable using conventional ladder prototypes --- either in reflection or transmission --- only now in transmission with no reflection whatsoever, from either port or at any frequency, given ideal elements. Many of these transfer functions were previously inaccessible by the recently developed reflectionless filter topologies as they would have required negative reactive elements, but the substitutions presented in this paper provide a way of eliminating those negative elements. Several test filters were made realizing Chebyshev, Butterworth, and Legendre transfer functions, with excellent agreement between measurement and theory.



\bibliographystyle{IEEEtran}
\bibliography{morgan}

\begin{thebibliography}{10}
\providecommand{\url}[1]{#1}
\csname url@samestyle\endcsname
\providecommand{\newblock}{\relax}
\providecommand{\bibinfo}[2]{#2}
\providecommand{\BIBentrySTDinterwordspacing}{\spaceskip=0pt\relax}
\providecommand{\BIBentryALTinterwordstretchfactor}{4}
\providecommand{\BIBentryALTinterwordspacing}{\spaceskip=\fontdimen2\font plus
\BIBentryALTinterwordstretchfactor\fontdimen3\font minus
  \fontdimen4\font\relax}
\providecommand{\BIBforeignlanguage}[2]{{%
\expandafter\ifx\csname l@#1\endcsname\relax
\typeout{** WARNING: IEEEtran.bst: No hyphenation pattern has been}%
\typeout{** loaded for the language `#1'. Using the pattern for}%
\typeout{** the default language instead.}%
\else
\language=\csname l@#1\endcsname
\fi
#2}}
\providecommand{\BIBdecl}{\relax}
\BIBdecl

\bibitem{AN-75-007}
\BIBentryALTinterwordspacing
{Mini-Circuits Inc.} (2015) Pairing mixers with reflectionless filters to
  improve system performance. [Online]. Available:
  \url{https://www.minicircuits.com/app/AN75-007.pdf}
\BIBentrySTDinterwordspacing

\bibitem{khalajamirhosseini2017}
M.~Khalaj-Amirhosseini and M.~M. Taskhiri, ``Twofold reflectionless filters of
  inverse-{Chebyshev} response with arbitrary attenuation,'' \emph{IEEE
  Transactions on Microwave Theory and Techniques}, vol.~PP, no.~99, pp. 1--5,
  2017.

\bibitem{psychogiou2017}
D.~Psychogiou and R.~G\'omez-Garc\'ia, ``Reflectionless adaptive {RF} filters:
  Bandpass, bandstop, and cascade designs,'' \emph{IEEE Transactions on
  Microwave Theory and Techniques}, vol.~65, no.~11, pp. 4593--4605, November
  2017.

\bibitem{chien2017}
S.-H. Chien and Y.-S. Lin, ``Novel wideband absorptive bandstop filters with
  good selectivity,'' \emph{IEEE Access}, vol.~5, pp. 18\,847--18\,861,
  Semptember 2017.

\bibitem{morgan_theoretical}
M.~A. Morgan and T.~A. Boyd, ``Theoretical and experimental study of a new
  class of reflectionless filter,'' \emph{IEEE Transactions on Microwave Theory
  and Techniques}, vol.~59, no.~5, pp. 1214--1221, May 2011.

\bibitem{morgan_structures}
------, ``Reflectionless filter structures,'' \emph{IEEE Transactions on
  Microwave Theory and Techniques}, vol.~63, no.~4, pp. 1263--1271, April 2015.

\bibitem{morgan8392495}
M.~A. Morgan, ``Reflectionless filters,'' U.S. Patent 8\,392\,495, March 5,
  2013.

\bibitem{morgan9705467}
------, ``Sub-network enhanced reflectionless filter topology,'' U.S. Patent
  9,705,467, July 11, 2017.

\bibitem{morgan9923540}
------, ``Transmission-line reflectionless filters,'' U.S. Patent 9,923,540,
  March 20, 2018.

\bibitem{morgan_optimal}
------, ``Optimal response reflectionless filters,'' U.S. Patent Application
  15/298,459, October 20, 2016.

\bibitem{morgan_deep}
------, ``Deep rejection reflectionless filters,'' U.S. Patent Application
  62/652,731, April 4, 2018.

\bibitem{morgan_artech}
------, \emph{Reflectionless Filters}.\hskip 1em plus 0.5em minus 0.4em\relax
  Norwood, MA: Artech House, 2017.

\bibitem{papoulis1958}
A.~Papoulis, ``Optimum filters with monotonic response,'' \emph{Proceedings of
  the IRE}, vol.~46, no.~3, pp. 606--609, March 1958.

\bibitem{bond2004}
\BIBentryALTinterwordspacing
C.~R. Bond. (2004, November) Optimal {``L''} filters: polynomials, poles, and
  circuit elements. [Online]. Available:
  \url{http://www.crbond.com/papers/optf2.pdf}
\BIBentrySTDinterwordspacing

\end{thebibliography}

\begin{IEEEbiography}[{\includegraphics[width=1in,height=1.25in,clip,keepaspectratio]{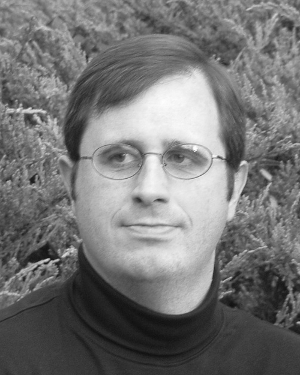}}]{Matthew A. Morgan}
(M'99--SM'17) received his B.S. in electrical engineering from the University of Virginia in 1999, and his M.S. and Ph.D. from the California Institute of Technology in 2001 and 2003, respectively.

During the summers of 1996 through 1998, he worked for Lockheed Martin Federal Systems in Manassas, VA, as an Associate Programmer, where he wrote code for acoustic signal processing, mathematical modeling, data simulation, and system performance monitoring. In 1999, he became an affiliate of NASA’s Jet Propulsion Laboratory in Pasadena, CA. There, he conducted research in the development of Monolithic Millimeter-wave Integrated Circuits (MMICs) and MMIC-based receiver components for atmospheric radiometers, laboratory instrumentation, and the deep-space communication network. In 2003, he joined the Central Development Lab (CDL) of the National Radio Astronomy Observatory (NRAO) in Charlottesville, VA, where he now holds the position of Scientist/Research Engineer. He is currently the head of the CDL’s Integrated Receiver Development program, and is involved in the design and development of low-noise receivers, components, and novel concepts for radio astronomy instrumentation in the cm-wave, mm-wave, and submm-wave frequency ranges. He has authored over 60 papers and holds seven patents in the areas of MMIC design, millimeter-wave system integration, and high-frequency packaging techniques. He is the author of \emph{Reflectionless Filters} (Norwood, MA: Artech House, 2017).

Dr. Morgan is a member of the International Union of Radio Science (URSI), Commission J: Radio Astronomy. He received a Topic Editor's Special Mention in the IEEE THz Transactions Best Paper competition and the Harold A. Wheeler Applications Paper Award in 2015.

\end{IEEEbiography}

\begin{IEEEbiography}[{\includegraphics[width=1in,height=1.25in,clip,keepaspectratio]{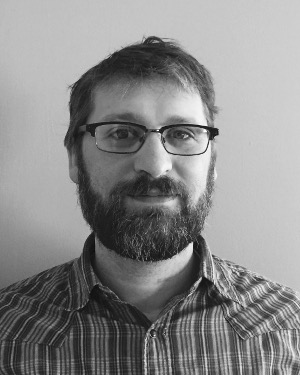}}]{Wavley M. Groves III} (S'11--M'16) received the A.A.S.E.T. degree from the Piedmont Community College, Charlottesville, in 2011.

From 2005 to 2007, he was a Professional Audio Repair Technician at Diversified Audio, Inc in Tampa, FL and manufactured avionics and ultrasound medical test equipment for UMA, Inc in Dayton, VA before joining the Low Noise Amplifier Group at the National Radio Astronomy Observatory (NRAO) as a Senior Technician in 2007.

Mr. Groves is currently a Technical Specialist II at NRAO and member of CDL's Integrated Receiver Development Program which he joined in 2012 where he is involved in the design, prototyping, and troubleshooting of prototype instrumentation refining it to become reliable radio astronomy solutions and is the Recording Engineer and Head Technician at EccoHollow Art \& Sound.
\end{IEEEbiography}


\begin{IEEEbiography}[{\includegraphics[width=1in,height=1.25in,clip,keepaspectratio]{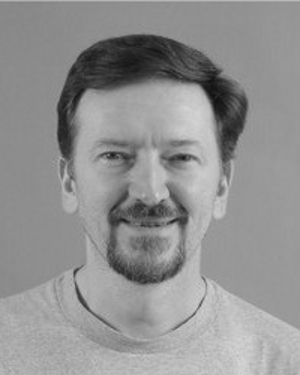}}]{Tod A. Boyd} was born in Steubenville, Ohio, in 1962. He received the A.S.E.E. degree from the Electronic Technology Institute, Cleveland, Ohio, in 1983.  From 1983 to 1985, he was with Hostel Electronics in Steubenville, Ohio.  In 1985, he joined Northrop Corporation’s Electronic Countermeasures Division, in Buffalo Grove, Ill., specialized in supporting the B-1B Lancer (secret clearance.)  In 1990, he joined Interferometrics, Inc., Vienna, Va., where he constructed VLBA tape recorders for the international Radio Astronomy community.

Since 1996, he has been with the National Radio Astronomy Observatory’s Central Development Lab, Charlottesville, VA, where he initially assisted with the construction of cooled InP HFET amplifiers for the NASA’s Wilkinson Microwave Anisotropy Probe (WMAP) mission.  Presently as a Technical Specialist IV he provides technical support for the advanced receiver R\&D initiatives. His responsibilities also include constructing low noise amplifiers for the Enhanced VLA and the Atacama Large Millimeter/sub-millimeter Array (ALMA) projects.
\end{IEEEbiography}




\end{document}